# THE EFFECT OF NON-THERMAL PROTONS ON THE HIGH ENERGY SPECTRA OF BLACK HOLE BINARIES

S. Bhattacharyya[1], N. Bhatt[1], R. Misra[2] and C. L. Kaul[1]

## ABSTRACT

In the inner regions of an accretion disk around a black hole, the presence of non-thermal protons would via p-p collisions produce electrons, positrons and $\gamma$-rays. For such a scenario, the steady state electron-positron distribution is computed taking into account Compton cooling, $e^-e^+$ pair production (due to $\gamma - \gamma$ interactions) and pair annihilation.

The resultant spectra has a broad feature around $1 - 10$ MeV which may be tested with observations by INTEGRAL and GLAST. Using the available OSSE data for the black hole system, GRS 1915+115 we illustrate that an upper limit on the non-thermal proton fraction can be obtained, which may put constraints on the acceleration process active in this system.

*Subject headings:* accretion, accretion disks—black hole physics

## 1. Introduction

Black hole X-ray binary systems are observed primarily in one of the two long term spectral states : the hard-state and the soft-state. In the so called hard state, the spectrum of the system can be described approximately as a hard power-law with a spectral index $\Gamma \approx 1.7$ and a cutoff around 100 keV, which suggest a thermal Comptonization origin (Shapiro *et al.* 1976; Liang & Price 1977). Indeed, detailed fits to broad band X-ray spectra confirm that the spectra is well represented by Comptonization of soft photons by a plasma having temperature $T \approx 50$ keV and optical depths of order unity (Gierlinski *et al.* 1997). In contrast, the soft spectral state consists of a black-body like component (kT $\approx 1$ keV) which generally dominates the luminosity, and a power-law tail with a photon index $\Gamma \approx 2.5$ with no detectable cutoff up to $\approx 600$ keV (Zdziarski *et al.* 2001). The steepness of the spectra and the absence of a spectral break indicates that this component arises due to Comptonization of soft photons by a non-thermal electron distribution. Indeed, the spectra can be well fit by a model where soft photons are Comptonized by a plasma consisting of both thermal and non-thermal electron distributions in a "hybrid" state (Zdziarski *et al.* 2001; Poutanen & Coppi 1998). Comptonization by thermal electrons ($kT \approx 1$ keV) gives rise to the soft component, while the power-law tail is produced by the non-thermal electrons. In this model, the steady state non-thermal electron distribution is computed self-consistently by assuming that there is an injection of non-thermal particles into the system where they are cooled by Comptonization and Coulomb interactions (Gierlinski *et al.* 1999). The salient results of this analysis suggests that the thermal and non-thermal components of the electron distribution co-exist in the same region and that the power in the non-thermal injection rates may be $\approx 10 - 20\%$ of the luminosity of the source.

[1]Nuclear Research Laboratory, Bhabha Atomic Research Centre, Mumbai-400085, India

[2]Inter-University Center for Astronomy and Astrophysics, Post Bag 4, Ganeshkhind, Pune-411007, India; rmisra@iucaa.ernet.in

The hard X-ray emitting region in the soft state can possibly be a corona or a flare on top of the cold accretion disk (Haardt & Maraschi 1993; Poutanen & Fabian 1999). While the energy released in a corona/flare is probably related to magnetic field reconnections, the details of the process are largely unknown. It is also not clear what kind of an electron acceleration mechanism (which is inferred from the non-thermal spectra) is active in these sources. If this mechanism is mass independent then one expects protons also to be accelerated, to form a non-thermal proton distribution. Protons may also be independently accelerated by scattering off magnetic "kinks" in a Keplerian disk (Subramanian *et al.* 1996). Thus evidence for a non-thermal proton distribution is expected to give valuable insight into the nature of the accelerating processes, and hence constrain the nature of the X-ray emitting region. Moreover, these protons may also escape from the inner regions and form relativistic jets (Subramanian *et al.* 1999) observed in some of these sources.

Relativistic protons can undergo inelastic collisions with thermal protons to produce charged and neutral pions which, in turn, decay into electrons, positrons and $\gamma$-rays. Since these electrons and positrons are primarily produced at high energies ($\gamma_e \approx 200$), their steady state distributions are expected to be steep (typically $\propto \gamma_e^2$). Thus their photon spectral signature will be distinct from the non-thermal electron distribution produced directly by the acceleration process (typically $\propto \gamma_e^{-2}$). Pair cascades initiated by injection of pairs in the system e.g. (Lightman & Zdziarski 1987; Svensson 1987) and pair production due $p - p$ reactions (Stern *et al.* 1992) have been studied extensively. In these works, Monte Carlo simulations have been undertaken to compute correctly the emergent spectra from a pair dominated plasma. Our motivation is to compute these spectral signatures for parameters pertaining to black hole binaries in the soft state which can be used to constrain the fraction of non-thermal protons in the system. We limit the study to plasmas which are not pair dominated which allows for approximations that make the analysis relatively simpler. An important effect, considered here, is the pair production of $\gamma$-ray photons in interaction with copious soft X-ray photons of energy $\sim 3$ keV that exist during the soft state. In this respect, this work differs from previous computation of spectra from non-thermal electron distributions due to pion decay , where pair production due to the presence of copious soft photons are not considered (Eilek & Kafatos 1983; Mahadevan *et al.* 1997; Markoff *et al.* 1999) and those where the soft radiation were assumed to be UV photons (Stern *et al.* 1992), scenarios more relevant to under-luminous black hole systems and AGN.

In the next section we describe the model and the assumptions made to compute the steady state non-thermal electron/positron distributions and the resultant photon spectra. In §3, generic results of the computation are presented along with a specific application to the black hole system GRS 1915+105. The work is summarized and main results are discussed in the last section.

## 2. Steady state electron/positron distribution and spectra

We consider a uniform sphere of non-relativistic thermal plasma with number density, $n_T$ and radius $R$. To represent the soft spectral component, it is assumed that inside the plasma, there is a Wein peak photon density $n_{\gamma,s}(e)$ at temperature $T_s$ (where $e \equiv h\nu/m_e c^2$ is the normalized photon energy) and normalization determined such that the photon energy density

$$U_{\gamma,s} = m_e c^2 \int_o^\infty e n_{\gamma,s}(e) de = L_s(1+\tau)/(c 4\pi R^2) \qquad (1)$$

Here $L_s$ is the luminosity of the soft component and $\tau = n_T \sigma_T R$ is the Thompson optical depth. Since the acceleration mechanism is unknown, we assume that a small fraction $f$ of the thermal protons form a non-thermal power-law distribution i.e

$$n_{NT}(\gamma_p) = f n_T \frac{\alpha - 1}{\gamma_{p,min}} (\frac{\gamma_p}{\gamma_{p,min}})^{-\alpha} \text{ for } \gamma_p > \gamma_{p,min} \qquad (2)$$

where $\gamma_p$ is the Lorentz factor of a proton and $\gamma_{p,min}$ is the minimum $\gamma_p$ of the distribution. It has been assumed here that the particle index $\alpha > 2$ and that the maximum Lorentz factor, $\gamma_{p,max} >> \gamma_{p,min}$.

These non-thermal protons interact with the thermal protons via $p - p$ reactions and subsequently produce electron/positron and $\gamma$-rays by pion decay. The threshold energy for such $p - p$ reactions is $\gamma_{p,thres} \approx 1.5$ while the other possible reactions ( e.g. $p - \gamma$ pair production) have a much higher threshold $\gamma_{p,thres} \approx 300$ and are neglected. It is not clear to what maximum Lorentz factor $\gamma_{p,max}$ would the unknown mechanism accelerate the protons. The absence of a spectral signature due to such reactions may simply indicate a lower value of $\gamma_{max}$ and not necessarily mean the absence of an efficient proton acceleration mechanism. Moreover, such spectral signatures would be sensitive to the proton power-law index, $\alpha$ assumed in this work, while the results obtained below are fairly independent of such details of the acceleration mechanism. Nevertheless for typical parameters used in this work, we *a posteriori* check the pair production rate due to the $p-\gamma$ reaction and confirm that it may be neglected.

The steady state positron particle density, $n_{e^+}(\gamma)$, is obtained by solving the kinetic equation

$$\frac{\partial}{\partial \gamma}[(\dot{\gamma}_c + \dot{\gamma}_{comp})n_{e^+}(\gamma)] + \dot{n}_{e^+}(\gamma) = Q_{+,\gamma\gamma}(\gamma) + Q_{+,pp}(\gamma) \qquad (3)$$

while the electron particle density, $n_{e^-}(\gamma)$, is obtained from

$$\frac{\partial}{\partial \gamma}[(\dot{\gamma}_c + \dot{\gamma}_{comp})n_{e^-}(\gamma)] = Q_{-,\gamma\gamma}(\gamma) + Q_{-,pp}(\gamma) \qquad (4)$$

Following (Eilek & Kafatos 1983), $e^+e^-$ production rate due to the $p-p$ process is given by

$$Q_{e^\pm,pp}(\gamma_e) = n_T c \int_{\gamma_{p,l}(\gamma_e)}^{\gamma_{p,h}(\gamma_e)} \frac{\sigma_{\pi^\pm}(\gamma_p) n_{NT}(\gamma_p) d\gamma_p}{[(\bar{\gamma}_\star - 1)(2\gamma_{pk}^{3/4} + \gamma_{pk}^{3/2})]^{1/2}} \qquad (5)$$

where

$$\gamma_{p,h}(\gamma_e) = 1 + [\bar{\gamma}_\star \gamma_e + (\bar{\gamma}_\star^2 - 1)^{1/2}(\gamma_e^2 - 1)^{1/2} - 1]^{4/3} \qquad (6)$$

$$\gamma_{p,l}(\gamma_e) = 1 + [\bar{\gamma}_\star \gamma_e - (\bar{\gamma}_\star^2 - 1)^{1/2}(\gamma_e^2 - 1)^{1/2} - 1]^{4/3} \qquad (7)$$

where $\bar{\gamma}_\star \equiv 70$. The approximate but analytical cross-section ($\sigma_{\pi^\pm}$) is tabulated for different energy ranges by (Eilek & Kafatos 1983)

Pair production rate from photon-photon (i.e. $\gamma\gamma$) interaction is approximated to be

$$Q_{\pm,\gamma\gamma}(\gamma_e) = c \int n_\gamma(2\gamma_e - e) n_\gamma(e) \sigma_{\gamma\gamma}(e, 2\gamma - e) de \qquad (8)$$

where it has been assumed that for two photons annihilating with energies $e$ and $e'$ the resultant $e^-$ or $e^+$ Lorentz factor is $\approx (e+e')/2$. The approximate form for the cross-section $\sigma_{\gamma\gamma}(e, e')$ is given by (Coppi & Blandford 1990). The positrons primarily annihilate with the background thermal electrons at a rate given by

$$\dot{n}_{e^+}(\gamma_{e^+}) = n_{e^+}(\gamma_{e^+}) n_T \sigma_{e^+e^-}(1, \gamma_{e^+}) c \qquad (9)$$

where the approximate form for the cross-section is used (Coppi & Blandford 1990). The annihilation of electrons has been neglected in eqn (4). $\dot{\gamma}_c$ and $\dot{\gamma}_{comp}$ are the rates of change of Lorentz factor due to Coulomb and inverse Compton cooling respectively. The inverse Compton cooling is primarily due to the up-scattering of soft ($\approx 3$ keV) photons to higher energies by non-thermal particles. Since in general the leptonic Lorentz factor may be $> 100$, the scattering in the rest frame may not take place in the Thompson limit and the general expression for the inverse Compton photon production rate and cooling (Jones 1968; Blumenthal & Gould 1970) have been used.

The equilibrium photon density inside the sphere is a solution of

$$Q_{\gamma,IC} + Q_{\gamma,pp} + Q_{\gamma,e^+e^-} = n_\gamma(e)[R_{\gamma\gamma} + \frac{c}{R(1+\tau_{KN}(e))}] \qquad (10)$$

where the rate of photon annihilation is given by

$$R_{\gamma\gamma}(e) = c \int n_\gamma(e')\sigma_{\gamma\gamma}(e,e')de' \qquad (11)$$

with the cross-section approximated to be (Coppi & Blandford 1990),

$$\sigma_{\gamma\gamma}(e,e') = \sigma_T \frac{(x-1)^{3/2}}{x^{5/2}}(\frac{1}{2}x^{-1/2} + \frac{3}{4}\ln(x))\mathrm{H}(x-1) \qquad (12)$$

$Q_{\gamma,IC}$, $Q_{\gamma,pp}$ and $Q_{\gamma,e^+e^-}$ are the photon production rates due to inverse Compton, $p-p$ interaction and pair annihilation, respectively. The latter is assumed to be

$$Q_{\gamma,e^+e^-}(e) = 4n_T n_{e^+}(2e-1)\sigma_{e^+e^-}(1,2e-1)c \qquad (13)$$

The photon production due to $p-p$ reaction is taken to be (Eilek & Kafatos 1983)

$$Q_{\gamma,pp}(e) = \frac{m_e}{m_\pi}n_T c \int_{\gamma_{p,l}(e)}^\infty \frac{\sigma_{\pi^0}(\gamma_p)n_{NT}(\gamma_p)d\gamma_p}{(\gamma_{pk}^{1/2} + 2\gamma_{pk}^{1/4})^{1/2}} \qquad (14)$$

where $\gamma_{pk} \equiv \gamma_p - 1$ and

$$\gamma_{p,l}(e_\nu) = 1 + [\frac{m_e e_\nu}{m_\pi} + \frac{m_\pi}{4m_e e_\nu} - 1]^{4/3} \qquad (15)$$

The approximate but analytical cross-section ($\sigma_{\pi^0}$) is tabulated for different energy ranges by Kafatos and Eilek (1983). The last term in eqn.(10) represents photons escaping from the system and $\tau_{KN}$ is the optical depth taking into account the Klien-Nishina cross-section.

Although the Thompson scattering depth $\tau$ of the thermal electrons could be of order unity, the corresponding Klien-Nishina optical depth for high energy ($> 1$ MeV) photons is small. Hence scattering of high energy photons with thermal electrons has been neglected. It should be emphasized that the above equations are valid only when the number density of non-thermal leptons is much smaller than the thermal one $n_T$. This assumption is violated when the plasma is pair dominated. It is known, that for some regions of parameter space, a pair runaway (or cascade) may be initiated and no steady state solutions for the leptonic density will exist. Solutions close in parameter space to this pair run away limit, may indeed be pair dominated and hence will not be accurately described by the above formalism. However, pair dominated plasma naturally exhibit a strong annihilation line at 511 keV which is not generally observed in black hole binaries.

Equations (10), (3) and (4) are solved iteratively to obtain the steady state particle densities and radiative flux as a function of six parameters: the Thompson optical depth $\tau$, the size $R$, the soft photon luminosity $L_s$, the soft photon temperature $T_s$, the fraction of non-thermal protons $f$ and the non-thermal proton index $\alpha$.

## 3. Results

The computed spectra for different values of the non-thermal proton fraction $f$ and for typical galactic black hole parameters are shown in Figure 1 while the corresponding steady state particle distributions are presented in Figure 2. The spectra are primarily due to inverse Comptonization of soft photons modified by $\gamma - \gamma$ absorption. Comparing the timescale for positron annihilation $t_{e^+} \approx 1/(n_T \sigma_T c)$ with that of inverse Compton cooling $t_{IC} \approx \gamma/\dot{\gamma}_{IC}$ reveals that

$$\frac{t_{e^+}}{t_{IC}} \approx \frac{n_{\gamma,s}}{n_T}\delta \approx 3 \times 10^3 \delta L_{s,38} R_7^{-1} T_{s,1}^{-1} \quad (16)$$

where $L_{s,38}$, $R_7$ and $T_{s,1}$ are the soft photon luminosity, the size and soft photon temperature normalized to $10^{38}$ ergs s$^{-1}$, $10^7$ cm and 1 keV respectively. $\delta \equiv \max(1, \frac{\gamma<e_s>}{m_e c^2})$ where $<e_s> \approx 3kT$ is the average soft photon energy. This implies that positron annihilation and the consequent annihilation emission is generally negligible compared to inverse Comptonization. In other words, positrons produced by $p-p$ or $\gamma-\gamma$ interactions cool by inverse Comptonization before annihilating. The luminosity due to $p-p$ interactions is

$$L_{p-p} \approx 8 \times 10^{36} \text{ergs s}^{-1} f\tau^2 R_7 \quad (17)$$

The high energy photons produced by the inverse Compton process, pair produce by interacting with the soft photons. This results in a spectral cut off around $\lesssim m_e^2 c^4/3kT_s \approx 80$ MeV. For high luminosities, the number density of $\gamma$-ray photons increases and pairs are also produced by self interaction of the $\gamma$-ray photons. The importance of this process may be characterized by the compactness parameter,

$$l_{p-p} \equiv \frac{L_{p-p}\sigma_T}{Rm_e c^3} \approx 20 f\tau^2 \quad (18)$$

such that when $l_{p-p} \gtrsim 1$, a luminosity dependent spectral break appears at an energy $> 1$ MeV. The three spectra in Figure 1, correspond to increasing $l_{p-p}$ of 0.8, 4 and 8 and show spectral breaks at different energies. The spectra with these breaks is qualitatively similar to that obtained in earlier works (Lightman & Zdziarski 1987) where the soft radiation was assumed to be UV photons.

To check *a posteriori* that other reactions apart from the $p-p$ is not important in such scenarios, we compute the pair production rate due to the $p-\gamma$ reaction using the photon and non-thermal proton densities corresponding to $f = 0.05$ and other parameters given in Figure 1. The reaction-rate expression for the processes was taken from (Chodorowski *et al.* 1992). The computed pair production rate $\approx 1.7 \times 10^{16}$ cm$^{-3}$/s is nearly two orders of magnitude smaller than that due to $p-p$ reaction $\approx 1.2 \times 10^{18}$ cm$^{-3}$/s. However, this result depends on the proton energy index $\alpha$ and for flatter distributions, $p-\gamma$ reactions may be important.

The resultant spectra generally have a broad feature around $1-50$ MeV and hence may be detected by INTEGRAL and from future observations by GLAST. Presently, OSSE data when combined with simultaneous low energy data, can provide upper limits on the non-thermal fraction $f$. As an illustration, we consider the case of GRS1915+105 which was observed in a soft state simultaneously by OSSE and RXTE on April 21 1999 (OSSE VP 813). (Zdziarski *et al.* 2001) fitted this data to the hybrid EQPAIR model and obtained the best fit black body temperature $= 1.35$ keV, the ion optical depth $\tau = 4.4$ and the luminosity of the soft photon source at a assumed distance of 12.5 kpc to be $1.3 \times 10^{39}$ ergs s$^{-1}$. On the assumption that these values may be approximated to be the soft photon temperature, optical depth and soft photon luminosity of this model, we compute the spectra for $R = 10^7$ cm, $\alpha = 2.5$ and for different values of $f$. Comparison of the observed OSSE flux at 500 keV ($EF_E \approx 0.1$ keV cm$^{-2}$ s$^{-1}$) with the computed ones, provides an upper limit on the non-thermal fraction $f < 0.05$. The corresponding limit on $p-p$ luminosity is $L_{p-p} < 8 \times 10^{36}$ ergs s$^{-1}$. On the other hand, the power injected in non-thermal electrons as inferred from the EQPAIR model fit is $\approx 5 \times 10^{37}$ ergs s$^{-1}$ (Zdziarski *et al.* 2001). Thus, on the basis of this simplistic and preliminary analysis, it would seem that the unknown accelerating mechanism preferentially energizes the electrons. However, it should be emphasized that an affirmative statement can only be made after a more careful analysis.

As mentioned earlier, the positron annihilation time-scale is longer than the inverse Compton one and hence the spectra shown in Figure 1 do not show any detectable annihilation lines. For low energy positrons $\gamma \approx 1$ the ratio of the time-scales is

$$\frac{t_{e+}}{t_{IC}} \approx 17\gamma L_{s,38} R_7^{-1} \qquad (19)$$

which is still larger than unity. Hence the positrons may cool and thermalize before annihi-lating and give rise to narrow ($\sigma \approx 1$keV) annihilation line. The high resolution spectroscopy of INTEGRAL may be able to detect such a line, which would be a self-consistency check and may put constraints on the size of the system. However, gravitational red-shift and broadening due to Keplerian motion of the disk may inhibit such a detection.

## 4. Summary and Conclusions

It is shown that the presence of non-thermal protons in the inner regions of an accretion disk around a black hole, may have a detectable high energy spectral signature. The non-thermal protons produce electrons, positrons and $\gamma$-rays by $p - p$ collisions. The ambient copious soft photons are Compton up-scattered by these pairs to $\gamma$-rays. Subsequently, these photons produce more $e^- e^+$ pairs and an equilibrium is reached with the steady state emergent spectrum having a broad feature around $1 - 50$ MeV. This spectral feature may be detected by future observations from INTEGRAL or GLAST. It is illustrated that presently available high energy observations of GRS 1915+105 by OSSE, impose an upper limit on the fraction of non-thermal protons to be $< 0.05$.

The detection of such a $p - p$ high energy spectral signature will open a new window on our understanding of the acceleration mechanism and environment around black hole systems. It would also warrant a more careful and sophisticated treatment of the spectral computation than what has been done here. In particular, the specific geometry of the plasma (e.g. spherical cloud, corona, flare) has to be considered. The effect of $p - p$ reactions between non-thermal protons and the cold disk ones need to be taken into account and a more realistic treatment of the soft photon source (which is probably the cold disk) needs to be undertaken. The escape of relativistic protons (perhaps along open magnetic field lines) may also need to be considered. As shown in this work, the non-detection of this spectral feature will put constraints on the fraction of non-thermal protons existing in a source. This in turn will put strong constraints on the acceleration mechanism, since that would mean that the unknown mechanism preferentially accelerates electrons.

The authors thank P. Subramanian for useful discussions.

– 9 –Eilek, J. A., & Kafatos, M., 1983, ApJ, 271, 804.

Gierlinski, M. *et al.*, 1997, MNRAS, 288, 958.

Gierlinski, M. *et al.*, 1999, MNRAS, 309, 496.

Haardt, F. & Maraschi, L. 1993, ApJ, 413, 507.

Jones, F. C., 1968, Phys. Rev., 167, 1159.

Liang, E. P., & Price, R.H. 1977, ApJ, 218, 247

Mahadevan, R. Narayan, R., & Krolik, J., 1997, ApJ, 486, 268.

Markoff, S., Melia, F., & Sarcevic, I., 1999, ApJ, 522, 870.

Poutanen, J., & Coppi, P.S., 1998, Phys. Scr., T77, 57.

Poutanen, J. & Fabian, A.C. 1999, MNRAS, 306, L31.

Shapiro, S. L., Lightman A. P. & Eardley, D. M. 1976, ApJ, 204, 187

Stern, B.E., Sikora, M. & Svensson R., 1992, Testing the AGN paradigm, Proceedings of the 2nd Annual Topical Astrophysics Conference, p. 313.

Subramanian, P., Becker, P. A., & Kafatos, M., 1996, ApJ, 469, 784.

Subramanian, P., Becker, P. A., & Kazanas, D., 1999, ApJ, 523, 203.

Svensson, R., 1987, MNRAS, 227, 403.

Lightman, A. P. & Zdziarski, A. A., 1987, ApJ, 319, 643.

Zdziarski, A. A. *et al.*, 2001, ApJ, 554, L45.

Chodorowski, M. J., Zdziarski, A. A. & Sikora, M., 1992, ApJ, 400, 181.
---

This preprint was prepared with the AAS LATEX macros v5.0.

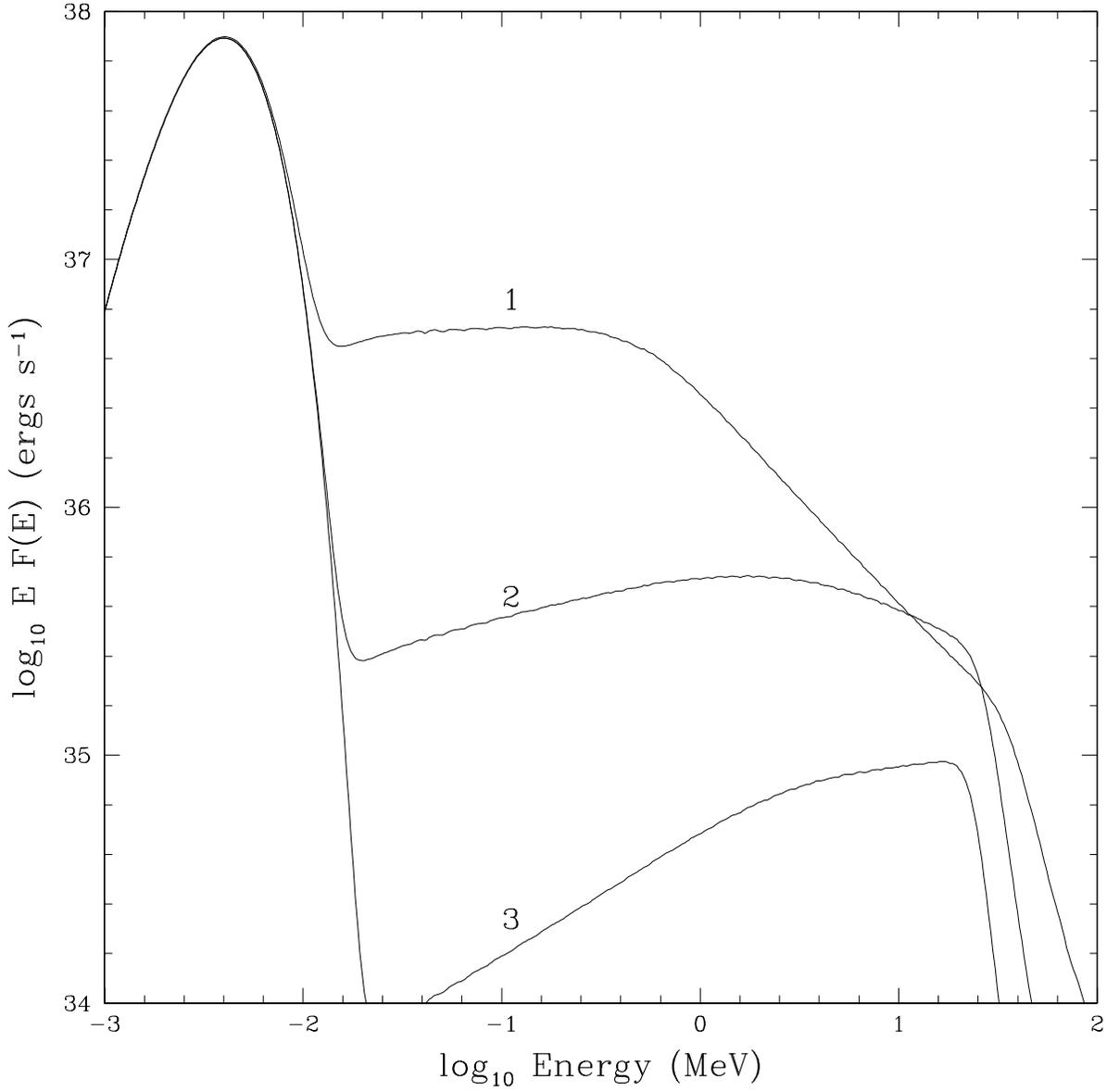

Fig. 1.— Computed spectra for three different values of the non-thermal proton $f = 0.1, 0.05, 0.01$ labeled as 1,2 and 3 respectively. The other parameters are $R = 10^7$ cm, $L_s = 10^{37}$ ergs s$^{-1}$, $kT_s = 1$ keV, $\tau = 2$ and $\alpha = 2.5$

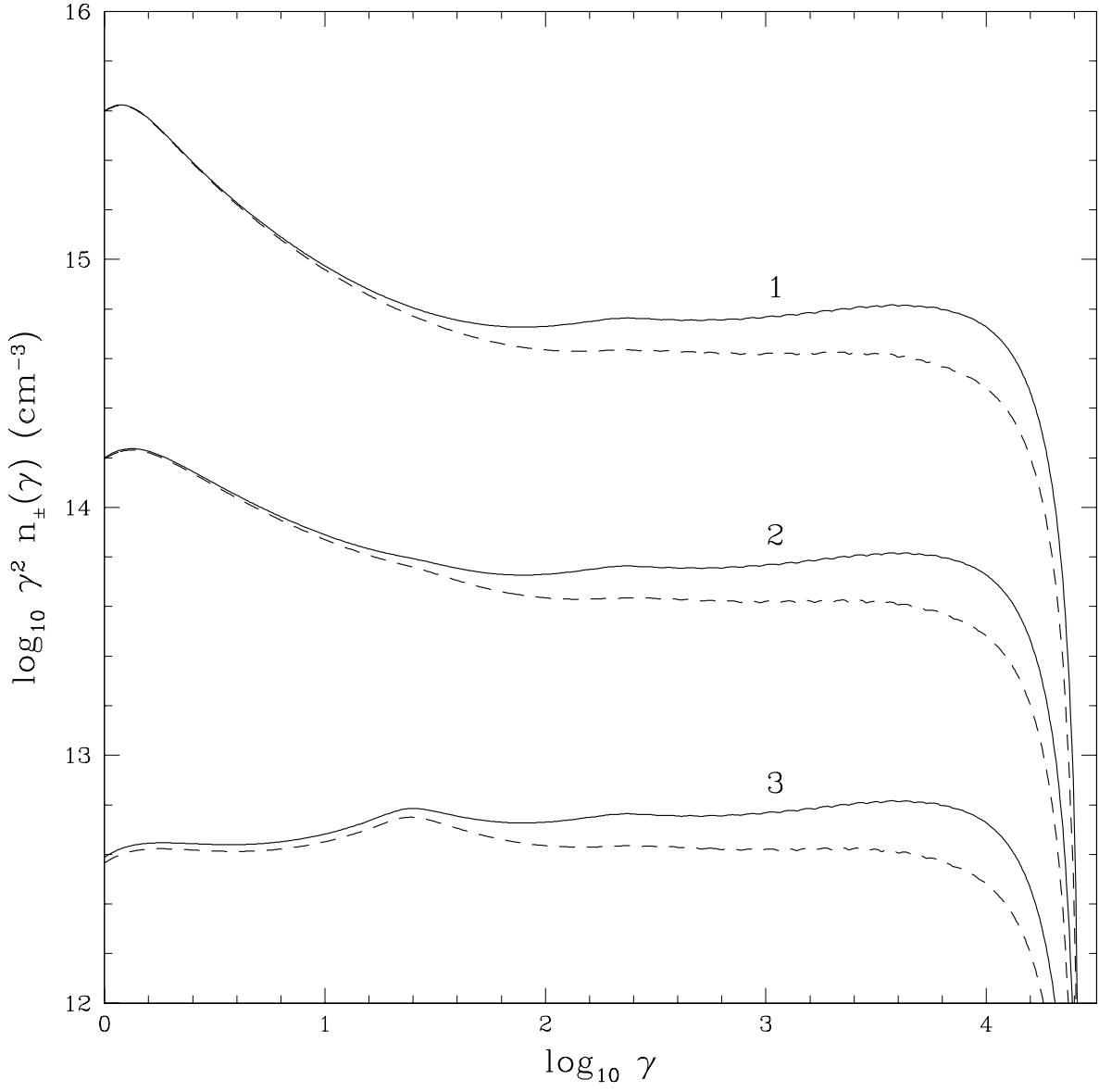

Fig. 2.— The steady state positron (bold) and electron (dashed) number densities in the plasma corresponding to the spectra in Figure 1